# Fiber-based Ratiometric Optical Thermometry with Silicon-Vacancy in Microdiamonds

Md Shakhawath Hossain,[†] Miguel Bacaoco,[‡] Thi Ngoc Anh Mai,[†] Guillaume Ponchon,[§] Chaohao Chen,[‖, #] Lei Ding,[††] Yongliang Chen,[‡‡] Evgeny Ekimov, [§§, ‖ ‖,] Helen Xu, [††] Alexander S. Solntsev,[‡] and Toan Trong Tran [†, *]

[†]School of Electrical and Data Engineering, University of Technology Sydney, Ultimo, NSW, 2007, Australia.

[‡]School of Mathematical and Physical Sciences, University of Technology Sydney, Ultimo, NSW, 2007, Australia.

[§]École polytechnique universitaire de l'université Paris-Saclay (Polytech Paris-Saclay), Université Paris-Saclay, Bâtiment 620, Maison de l'Ingénieur, Rue Louis de Broglie, 91190 Orsay.

[‖]Department of Electronic Materials Engineering, Research School of Physics, The Australian National University, Canberra, Australian Capital Territory 2601, Australia.

[#]ARC Centre of Excellence for Transformative Meta-Optical Systems (TMOS), Research School of Physics, The Australian National University, Canberra, Australian Capital Territory 2601, Australia.

[††]School of Biomedical Engineering, University of Technology Sydney, Ultimo, NSW, 2007, Australia.

[‡‡] Department of Physics, The University of Hong Kong, Pokfulam, Hong Kong, China.

[§§]Institute for High Pressure Physics, Russian Academy of Sciences, Troitsk 142190, Russia.

[‖ ‖]Lebedev Physics Institute, Russian Academy of Sciences, Moscow 117924, Russia.

*Corresponding author: trongtoan.tran@uts.edu.au

## ABSTRACT

Fiber optic all-optical thermometry is a promising technology to track temperature at a micro-scale while designing efficient and reliable microelectronic devices and components. In this work, we demonstrate a novel real-time ratiometric fiber optic thermometry technique based on silicon-vacancy (SiV) diamond that shows the highest temperature resolution (22.91 $KHz^{-\frac{1}{2}}Wcm^{-2}$) and spatial resolution (~7.5 µm) among all-optical fiber-based thermosensors reported to date. Instead of analyzing the spectral features of temperature-dependent SiV signal, coming from SiV micro-diamond fixed on the fiber tip, an alternative parallel detection method based on filtering optics and photon counters is proposed to read out the sample temperature in real-time. The signal collection efficiency of the fiber is also investigated numerically with semi-analytic ray-optical analysis and then compared with our experimental study. We finally demonstrate the performance of the thermosensor by monitoring the temperature at distinct locations in a lab-built graphite-based microheater device. Our work introduces a reconfigurable method for temperature monitoring in microelectronic, microfluidic devices, or biological environments and unlocks a new direction for fiber-based all-optical thermometry research.

Current technological advancements in micro or nanoelectronics,[1-3] nanophotonics,[4, 5] biophotonics,[6] microfluidics,[7, 8] and nanomedicine[9] have reached a point where conventional thermometry is no longer applicable for measuring the temperature in the micro-scale limit due to the drop of spatial resolution to the micron level. To design reliable and efficient microelectronic devices and components, where their operational performance is often constrained by efficient heat dissipation, it is necessary to measure temperature precisely at the micro-scale. This imperative arises as thermal hot spots may arise suddenly during operation and cause performance deterioration, overheating, or even irreversible failure of these devices.[10] In recent years, optical thermometry has become popular among researchers due to its ability to monitor temperature changes remotely with high sensitivity and resolution in miniaturized microelectronic devices.[11] Many optical thermometry techniques based on different particles such as organic dyes,[12] quantum dots,[13] up-conversion nanoparticles,[14] hBN flakes,[10] and diamond[15] have been reported until now based on improving various properties of particles or completely discovering new materials that work on different target-specific temperature ranges.

Among these materials, diamond is gaining tremendous interest in all-optical thermometry due to its thermal stability, high thermal conductivity, compatibility with harsh and extreme conditions, and biocompatibility.[16] Diamond can host color centers—atom-like defect complexes that emit single photons. Nitrogen-vacancy (NV) color centers and group IV color centers such as germanium-vacancy (GeV), silicon-vacancy (SiV), lead-vacancy (PbV), tin-vacancy (SnV), etc. are some of the most common diamond color centers, owing to their compelling photophysical properties.[17] Temperature changes alter various spectral characteristics of diamond color centers such as zero phonon line (ZPL) wavelength,[18] ZPL linewidth,[19] photoluminescence (PL) intensity,[20] shift in optically detected magnetic resonance (ODMR) spectra,[21] etc. which are then analyzed to set up the correlation between the observable and temperature. However, most of these methods require post-processing of the acquired data or take a longer time to acquire spectral features which can lead to slower temperature measurements. Recently, this limitation has been alleviated by a high-speed ratiometric all-optical thermometry method which can monitor temperature changes in real-time of a microcircuit.[22] However, the lack of reconfigurability and portability are the major drawbacks of these free-space optical thermometers. Once the diamond thermometers are deposited onto a target device, it is very challenging to relocate them to different locations of the same device or recover and reuse them on another device, owing to the strong thermometer-sample interaction and the minuscule size of the thermometers. One way to circumvent this issue is to attach the optical thermometer to the tip of an optical fiber so the thermometer can be easily retrieved after a measurement is taken. Furthermore, fiber-based optical thermometry offers alignment-free between the optical beam and the thermometer, thanks to the physical attachment of the thermometer onto the center of the fiber tip. Such a feature allows fiber-based optical thermometers to be operated in a more reliable and robust manner without the risk of the thermometer moving out of the focused laser beam due to thermal or mechanical drifting. Compared to other traditional thermometers such as micro-thermocouples or micro-thermistors, fiber-based optical thermometers are more robust against harsh environments and are immune to electromagnetic interference—making them a favorable candidate for a variety of industrial applications.[23] There have been, however, only a few reports on fiber-based diamond thermometry to date,[24-27] in part due to the challenging integration of the optical thermometer onto the fiber facet. Most studies in the literature employed the NV center in diamond as the optical thermometer—owing to its optically accessible spin—which can be initiated, manipulated, and read out via the combination of optical and microwave excitation. Although the thermometer features an extremely impressive sensitivity,[28, 29] it requires the generation of microwave radiation for its operation. Such microwave radiation may induce heating which may not be discernible from heating caused by temperature changes and also makes the system semi-optical.[30] All-optical fiber-based thermometry that only uses optical excitation is, therefore, an excellent solution to this problem. Recently, an all-optical fiber-based thermometry method with GeV diamond has been reported with a spatial resolution of 25 μm, albeit with some drawbacks.[27] In this work, we propose an all-optical fiber-based diamond thermometry technique that addresses these limitations. The motivation of this work is threefold. The first is fabricating a reconfigurable, flexible thermal sensor alleviating the need for reproduction of new thermosensors for new samples/devices. The second is investigating the signal collection efficiency of the sensor with ray-optical analysis and comparing it with our experimental results. The third is using this thermal sensor to monitor the real-time temperature of microelectronic devices in operating conditions which could provide insightful information to researchers in the field of microelectronics, aiding them in the more efficient and optimized design of such devices.

In this work, we fabricate a fiber-based thermal sensor by deterministically depositing the SiV microdiamond on the facet of a multimode fiber core. By using the photoluminescence ratio of the SiV emission and the backscattered laser, instead of the conventional spectral analysis, we evince a record-high temperature resolution relative to excitation power of $22.91 \text{ KHz}^{-\frac{1}{2}}\text{Wcm}^{-2}$ and an excellent spatial resolution of 7.5 μm. We successfully apply the technique to measure local temperature on a lab-built graphite-based microheater device.

## RESULTS AND DISCUSSION

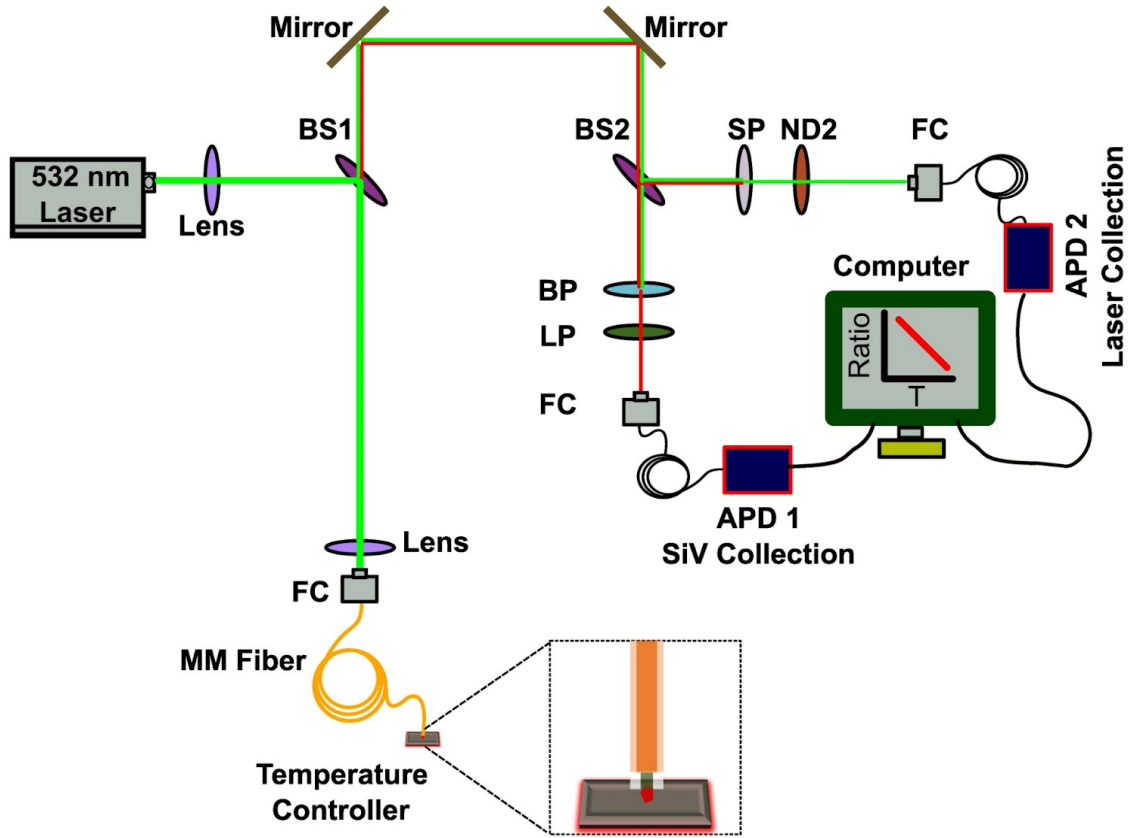

**Figure 1**. Schematic of the proposed fiber-based optical thermometry setup. A 200 μm core multimode fiber (MM fiber), flat cleaved end glued with SiV diamond, makes contact with the heating stage of a temperature controller. The other end of the fiber coupled with the fiber coupler (FC) is excited with a green laser (532 nm, 10 mW power) which is reflected from a 30R/70T beam splitter (BS1). The fluorescence SiV signal along with the laser signal from the multi-mode fiber bare end attached with SiV diamond is first transmitted through a 30R/70T beam splitter (BS1), guided through mirrors, and then split by another 50R/50T beam splitter (BS2). The SiV and laser signals are then filtered and collected by two avalanche photodiodes (APDs) on separate arms. For SiV signal collection, long-pass (LP) (715 nm) and band-pass (BP) filters (747/33 nm) are used. For laser signal collection short-pass (SP) filter (532 nm), and neutral density (ND2) filters are used.

We first describe the experimental setup used in this work. **Figure 1** illustrates the schematic of the lab setup **(cf. Methods)** to acquire the calibration curve. The bare fiber end attached to the SiV diamond is placed on the heating stage of a high-precision temperature controller. A 532 nm green continuous-wave (CW) laser (green line) excites the SiV diamond emitter attached to the flat cleaved end of the fiber through the other end of the fiber, producing

~739 nm SiV emission (red line). The emitted signal is first transmitted through a 30R/70T beam splitter and then split by another 50R/50T beam splitter. The split signals in two collection paths are collected by two fiber-coupled avalanche photodiodes (APDs). One APD collects the filtered SiV signal (red line) and the other APD collects the backscattered laser signal (green line) depending on their filter combination. The long-pass (LP) 715 nm and band-pass (BP) 747/33 nm filters in the SiV signal collection path block the backscattered laser signal and maximize the SiV count collected by APD1. On the other hand, in the laser signal collection path, the short-pass (SP) 532 nm filter blocks the SiV signal and the ND2 filter reduces the laser counts that APD2 collects, protecting APD2 from saturation. The SiV photoluminescence intensity varies with temperature while the laser intensity remains temperature-invariant and constant. The ratio of two APD signals is directly recorded in a computer display and can be converted to temperature data using a calibration curve (see below). Our real-time ratiometric method eliminates the need for post-processing of acquired emission spectra, thereby making the system response significantly faster.

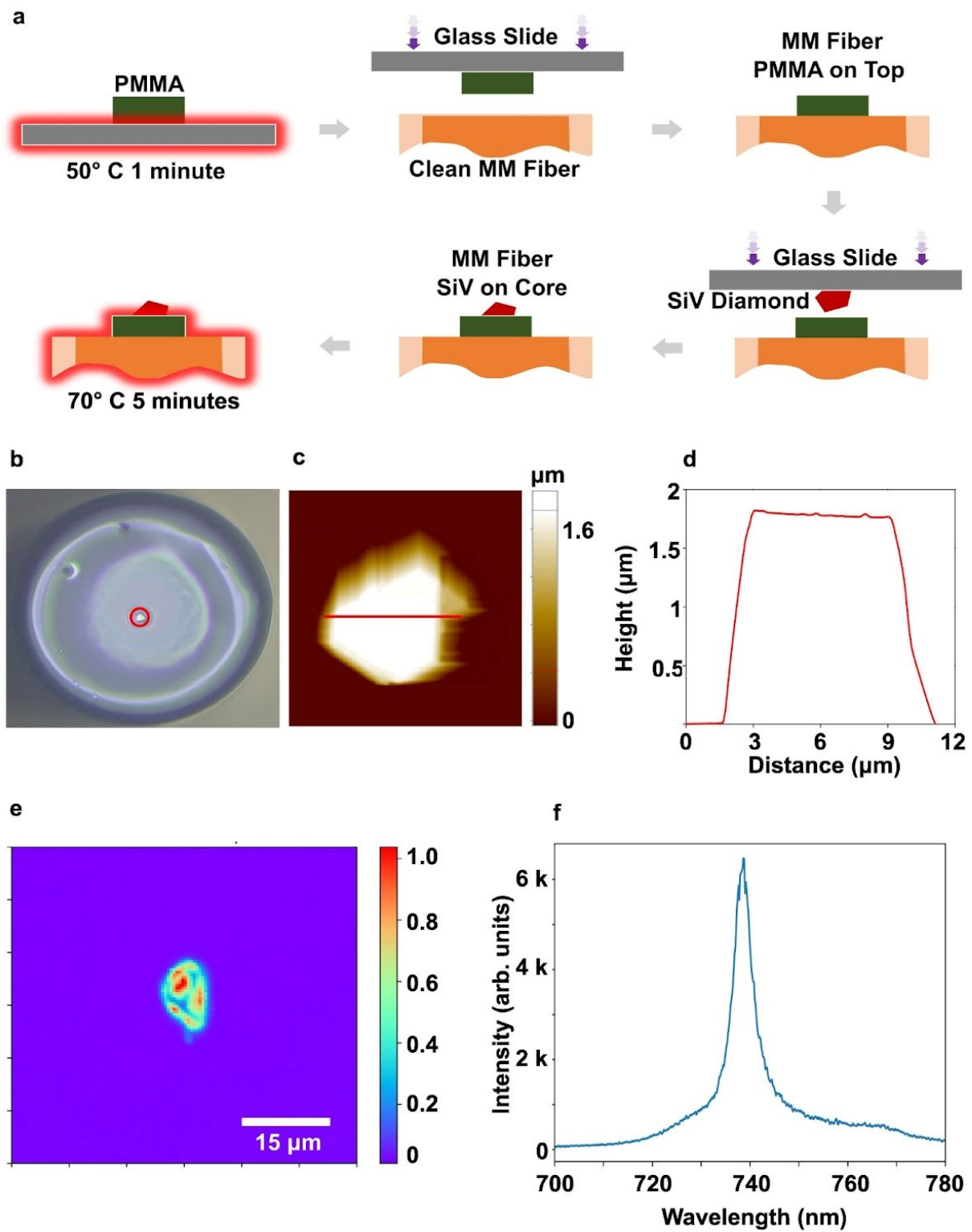

**Figure 2** Deterministic transfer and characterization of the transferred SiV diamond on the multimode fiber. (a) Schematic of the developed transfer process (see main text) to transfer any selected SiV diamond deterministically to any targeted position of the multimode fiber core. (b) Optical image of the transferred SiV diamond (marked by the red circle) on the multimode fiber core. (c) The corresponding atomic force microscopy (AFM) image of the SiV diamond. (d) The cross-sectional profile of the SiV diamond. The size of the SiV diamond is ~7.5 μm and the cross-section data refer to the area specified by the red line. (e) Confocal image of the transferred SiV diamond under excitation of 300 μW, 532 nm green laser through the objective. (f) Photoluminescence spectrum of the SiV diamond on the multimode fiber after transfer.

The active temperature-sensing element in our work is the SiV microdiamond mounted onto the fiber facet. To enable the precise transfer of the SiV micro-diamond, we adapt the all-dry transfer method previously developed elsewhere for transferring two-dimensional materials,[31] albeit with some modifications. **Figure 2a** demonstrates the procedure of transferring a single SiV diamond precisely onto a specific position of a multimode fiber core. First, a multi-mode fiber with a core diameter of 200 μm is cleaned with fiber cleaner that is soaked with acetone and isopropanol (IPA). A small drop of PMMA (polymethyl methacrylate) is drop cast on a glass slide and heated up to 50° C for 1 minute. PMMA on the glass slide then comes into contact with the fiber core with the lab-built XYZ stamping stage. The PMMA layer on the fiber core acts like glue to attach the diamond that is deposited onto it in the following step. A small portion of the pre-synthesized (**cf. Methods**) SiV diamond solution is drop-cast on a clean glass slide and left to dry for a few minutes. Next, to transfer the SiV diamond onto the fiber core, the fiber is mounted with a fiber chuck onto the sample stage and the glass slide stamp with the diamond is mounted upside down onto the stamping stage of our lab-built transfer setup (**cf. Methods**). The targeted SiV diamond on the glass slide is aligned to the fiber core using the zoom lens imaging system of the transfer setup. The transparency of the glass slide allows us to align the targeted SiV diamond to any position of the fiber core with our imaging system. The glass slide is slowly lowered until the targeted diamond makes contact with the selected position of the fiber core. After waiting for ~2 minutes to allow for maximum adhesion, the glass slide is slowly lifted using the Z micro-positioner, and the process is continuously monitored in our imaging system. Finally, the SiV diamond is transferred onto the specific location of the fiber core and is observed in our imaging system. The fiber core with SiV diamond is then heated up to 70° C for 5 minutes to cure the PMMA and thus lock the diamond in place. **Figure 2b** shows the wide-field optical image of the fiber after transferring the SiV diamond. As illustrated in **Supporting Information Figure S1**, a narrow diamond peak is observed at ~1335 cm$^{-1}$ with a full-width half-maximum (FWHM) of ~2.5 cm$^{-1}$ after transfer confirmed by Raman spectroscopy. Such Raman shift values imply the high crystallinity of the diamond after the transfer process. To characterize the size of the diamond, we employ atomic force microscopy (AFM). **Figure 2c** illustrates the AFM image of the corresponding SiV diamond and **Figure 2d** shows the height cross-section analysis of the AFM image. The horizontal and vertical dimensions of the SiV diamond are 7.5 and 1.8 µm, respectively. In fiber-based optical thermometry, the probe size determines the spatial resolution of the system, which is 7.5 µm, in our case. We perform confocal fluorescence mapping to characterize the SiV emission distribution on the microdiamond probe. **Figure 2e** shows the confocal maps of a 60 µm × 60 µm of the fiber core showing very bright spots from the diamond containing an ensemble of SiV emitters. The non-uniform distribution of these bright spots within the micro-diamond can be attributed to the high refractive index of the diamond. **Figure 2f** illustrates the SiV signal intensity collected by the spectrometer when the diamond on the fiber is excited with the 300 µW laser power (1s acquisition time). Here, diamond emitters are excited through the objective, and the emitted signal is also collected by the objective and guided in free space before finally being displayed on the spectrometer (**cf. Methods**).

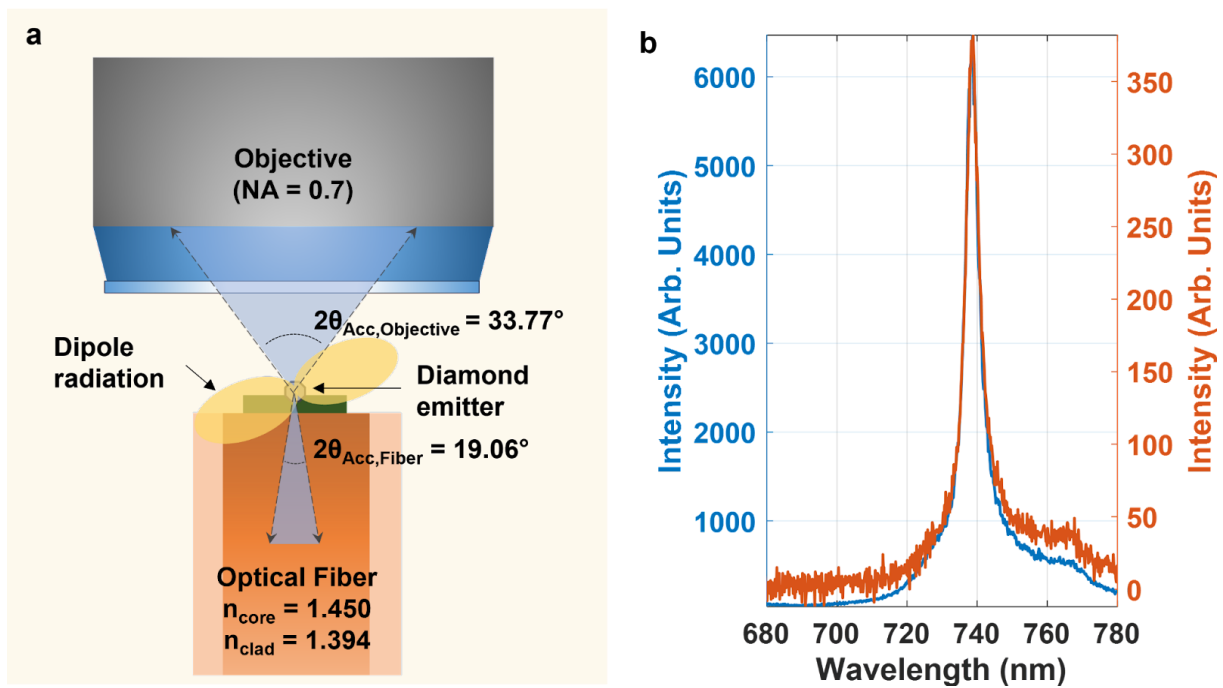

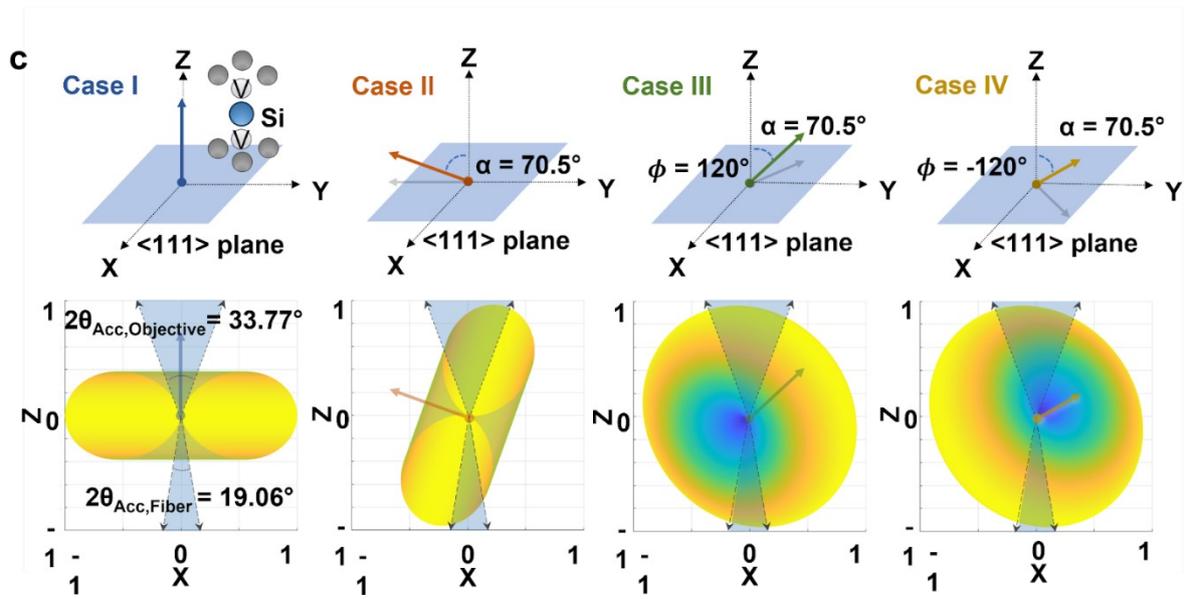

**Figure 3.** An analysis of experimental data and semi-analytical ray optics data on the collection efficiency of the SiV signal from the multimode fiber. (a) Schematic representation of the mechanism for determining the collection efficiency of the SiV diamond emitters through an optical fiber and objective lens. The PL emission of the diamond in the form of dipole radiation is collected by the same objective and by the optical fiber on the top and bottom direction of the emitter, respectively. The fraction of emission that gets coupled through the fiber and objective is then calculated from their respective acceptance angles ($2\theta_{Acc, Objective} = 33.77°$, $2\theta_{Acc, Fiber} = 19.06°$). (b) Emission spectra from the SiV ensemble collected by the optical fiber (orange) and objective (blue). Comparing the integrated intensities of the experimental results shows that optical fiber-based collection corresponds to 5.546% that of objective-based collection. (c) Due to the $D_{3d}$ point group symmetry of SiV diamond, four possible dipole orientations can manifest in the material system herein denoted as Cases I, II, III, and IV. The emission profile per dipole case is illustrated and their overlap to the acceptance cones defined by the acceptance angle centered along the z-axis is graphically shown. For Case I, negligible coupling is expected due to weak overlap of the dipole radiation to the acceptance cone: 0.095% and 0.530% of the total emitted radiation for fiber and objective, respectively. (**cf. Methods**). Most of the accepted radiation comes from oblique dipole orientations (Cases II, III, IV) where the overlap is greater. The oblique cases are symmetric along the z-axis and thus each contributes an equal amount of accepted radiation: 0.824% for fiber and 2.558% for the objective. Overall, the total accepted radiation from all orientations is calculated to be 0.642% for fiber and 2.051% for the objective. The simulation therefore predicts a 31.3% collection efficiency for the fiber in comparison to objective-based free space collection.

The collection efficiency of the optical fiber with respect to the free-space collection via the objective is investigated by determining the portion of the SiV diamond PL emission collected through the apertures of the fiber and the objective, respectively. **Figure 3a** shows the schematic representation of the simultaneous collection of light from the emitter to the objective and optical fiber. The resulting spectral measurement shown in **Figure 3b** shows no difference in line shape and peak location between the two collection methods. This suggests the robustness of fiber-based collection in spectral measurement is at par with free-space collection with no spectral distortion. However, a decrease in intensity is observed as expected from the difference in their numerical apertures, with the objective (NA=0.7) having higher light collection capacity than the optical fiber (NA=0.39). The measurement shows 5.546% collection efficiency for fiber-based collection with respect to the free-space objective-based collection.

To investigate the difference in intensities from the spectral measurement, a semi-analytic ray-optical analysis is then performed, considering the far-field spatial radiation pattern of the SiV emission and the acceptance angles of optical fiber and objective. In free space optics, the acceptance angle ($\theta_{Acc}$) is defined by the maximum angle from the optical axis allowed for a light ray to be transmitted through an aperture characterized by its numerical aperture (NA).[32] The acceptance angle ($\theta_{Acc}$) can be derived from the numerical aperture (NA) via equation (1), where $n_{medium}$ is the index of refraction of diamond ($n_{diamond}$=2.41) where the light is propagating from. This gives the acceptance angle of the objective ($\theta_{Acc, Objective}$) within the diamond slab equal to 16.88°.

$$\theta_{Acc,Objective} = \arcsin\left(\frac{NA}{n_{diamond}}\right) \quad (1)$$

The acceptance angle has an analogous definition in fiber optics where it is defined as the maximum angle for a light ray hitting the fiber core to be guided (via total internal reflection) within the fiber core.[32] For simple single-step multimode fibers, it is determined by the indices of refraction of the cladding ($n_{clad}$=1.394), core ($n_{core}$=1.450), and the medium where the light is incident from ($n_{diamond}$=2.41) shown in equation (2). Thus, the acceptance angle of the fiber ($\theta_{Acc, Fiber}$) within the diamond is 9.53°.

$$\theta_{Acc,Fiber} = \arcsin\left(\frac{1}{n_{diamond}}\sqrt{n_{core}^2 - n_{clad}^2}\right) \quad (2)$$

SiV diamond possesses a D3d point group symmetry that results in four possible dipole orientations within the material system shown in **Figure 3c**.[33] The dipole orientations denoted as Cases I, II, III, and IV, correspond to its dipole axis orthogonal to <111>, <**1**11>, <1**1**1>, and <11**1**> crystallographic planes, respectively. Each dipole is treated as a dipole emitter with far-field radial intensity distribution proportional to $\sin^2(\alpha)$, where $\alpha \in [0, \pi]$ is the altitude angle from the dipole axis.[34] Thus with this model, the radiation is axisymmetric along the dipole axis, and maximum emission is expected along the orthogonal plane ($\alpha = 90$). For each dipole case, the overlap of the emission profile to the acceptance cone defined by the acceptance angles of the fiber and objective is determined, and the total accepted emission is quantified as the sum of the accepted radial distances. Similarly, the total emission per dipole case is calculated as the sum of all radial distances of the dipole radiation. In the analysis, an ensemble of these dipole moments composes the emitter sample with presumed equal probabilities. Thus, the accepted emission per case is summed with equal weights and the theoretical collection efficiency is evaluated as the ratio of the total accepted light with respect to the total emission.

**Supporting Information Table S1** tabulates the calculated collection efficiency per dipole case. Due to the dipole-type radiation pattern of the emitter, minimal coupling is expected for Case I where the high-emission plane is at the right angle to the apertures (See **Figure 3c**). This is supported by the simulation results showing 0.095% and 0.530% collection efficiency for fiber and objective, respectively. On the other hand, higher coupling efficiency is expected for oblique dipoles (Cases II, III, IV) due to improved alignment of the dipole high-emission plane to the apertures. Collection efficiency for each oblique dipole was calculated to be 0.824% for the optical fiber and 2.558% for the objective. The oblique cases are $C_3$ rotationally symmetric along the z-axis, and thus each contributed an equal amount of accepted radiation. Summing up the contribution from all independent emission profiles, the simulation predicts a $\frac{0.642\%}{2.051\%} = 31.294\%$ efficiency for optical fiber-based collection with respect to the free-space objective-based collection. This result sets the theoretical upper limit to the collection efficiency in the absence of fabrication defects and optical alignment tolerances. However, in practice, unavoidable material and measurement imperfections may exist which could contribute to the experimental efficiency of 5.545%. Nevertheless, with such efficiency, the proposed fiber-based optical thermometric system can demonstrate a novel functionality with record high temperature resolution (see below) using off-the-shelf optical cable and without sophisticated collection optics.

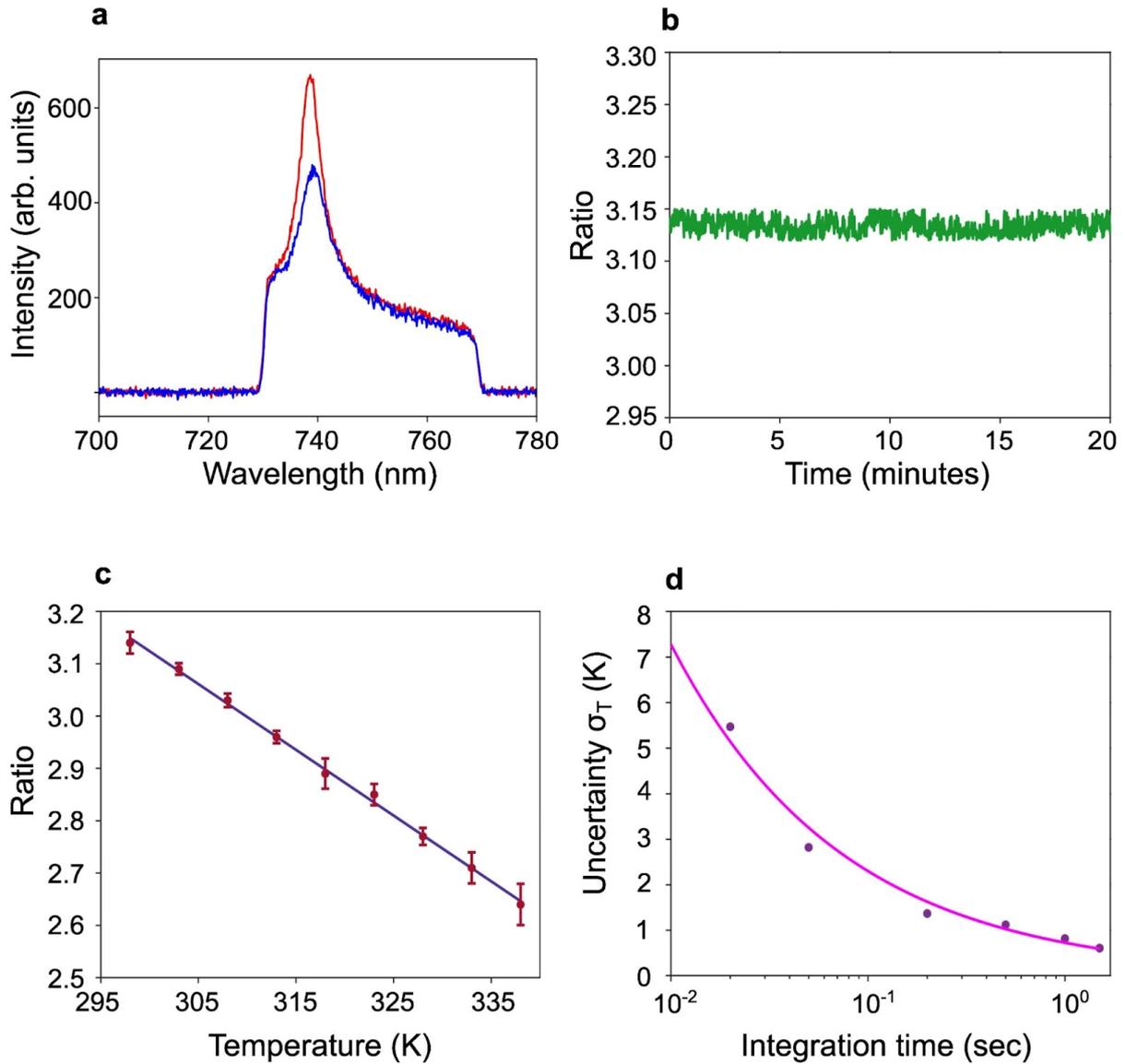

**Fig. 4** Fiber-based optical thermometry based on SiV/laser count ratio. (a) Photoluminescence spectra of the SiV ensemble at different temperature (red spectra indicates PL intensity at 298 K and blue spectra indicates PL intensity at 338 K). As temperature increases, the PL intensity of the SiV diamond attached to the fiber tip decreases. Each spectrum is collected with a 532 nm green laser (20 s acquisition time, 10 mW excitation power). (b) Stability measurement of SiV/laser count at room temperature (298 K) for 20 minutes. The ratio is recorded at 1s integration time. (c) Ratio of SiV to laser counts as the function of temperature. Each data point at a specific temperature is computed by taking the average of the data for 200 s, with 1s integration time. The indigo color line is a linear fit of the data points which decreases linearly by increasing temperature. The error bar of each data point is measured by taking the standard deviation of all the ratio values computed at the specific temperature and is marked by the maroon color. (d) Uncertainty of temperature ($\sigma_T$) versus integration time of the fiber optic diamond thermometer. Temperature resolution is defined by the equation $\eta_T = \sigma_T\sqrt{t_m}$, where $t_m$ is the integration time. It can be calculated from the shot noise fit (purple line). Each uncertainty data point ($\sigma_T$) is calculated by taking the average of 200 data points at every single integration time. All the ratio data from two APDs are taken using a 10 mW, 532 nm green laser.

Next, the performance of our proposed fiber optic diamond thermometer is tested. However, instead of exciting the diamond emitter attached to the fiber through the laser, the diamond emitter is excited through the other end of the optical fiber, and the signal is also collected from the same fiber port which is then delivered to the detector (cf. **Methods**). The diamond probe on the fiber core contacts the heating stage equipped with a temperature controller. The temperature is varied from 298 K (25° C) to 338 K (65° C) and the PL intensity of SiV at the two temperatures is observed on the spectrometer. The PL intensity of SiV decreases as the temperature increases as observed in **Figure 4a**. The reduction of this PL intensity of SiV is mainly due to the increased phonon activity at high temperatures that leads to the nonradiative energy transition.[35] However, the laser intensity remains constant as temperature increases (**Supporting Information Figure S2**). As a result, the intensity ratio of SiV/laser decreases as the temperature rises. To verify the stability of the intensity ratio, the ratio data is taken at room temperature (298 K) for about 20 minutes as illustrated in **Figure 4b**. The ratio fluctuates a little, with a value close to 0.0077, indicating an excellent long-term stability of our thermometer. **Figure 4c** depicts the plot of the SiV/laser intensity ratio as a function of temperature when the temperature is varied from 298 K to 338 K, with a 5 K increment. The data are fitted well with a linear function $f(T) = -mT + C$, where $f(T)$ is the fitting function, m is the slope of the equation, $T$ is the temperature, and $C$ is the offset constant. The absolute sensitivity can be calculated from the fit equation which is 1.25% K$^{-1}$. From the fit, we also have calculated the relative sensitivity $S_r = \frac{1}{O}\frac{dO}{dT}$, where $O$ is the measured observable, of 0.47 % K$^{-1}$ at 338 K and 0.39 % K$^{-1}$ at 298 K. Furthermore, the repeatability of the calibration curve is tested by plotting the ratio data for two consecutive heating and cooling cycles as shown in **Supporting Information Figure S3**. The overlapping of the calibration curve in different heating and cooling cycles indicates the robustness of the sensor. In addition, the ratio also remains constant for different laser excitation power as demonstrated in **Supporting Information Figure S4**. Since our technique relies on the SiV/laser ratio, the backscattered laser intensity plays an important role in determining the accuracy of the temperature readouts. Surfaces with different reflectivity of the excitation laser (532 nm) will result in some absolute changes in both the SiV emission and backscattered laser intensity, and thus the SiV/laser ratio. The laser reflectivity is, however, temperature independent as shown in **Supporting Information Figure S5,** where the ratio remains the same at different temperatures. In this reference test, the fiber-based sensor is separated from the temperature-controlled stage by 1 cm and temperature is varied from 298 K to 338 K with 5 K increments and hence there is no heat exchange between the sensor and the stage. In order to circumvent the inconsistencies among different sample surfaces, we first measure the SiV/laser ratio of the new surface at room temperature. By subtracting the ratio taken from the new surface to the standard surface (stainless steel of the temperature controller in this case) at room temperature, we arrive at the offset value—which is invariant with temperature. We then introduce this offset directly to the temperature calibration curve previously obtained from the standard surface (in **Figure 4c**). In this way, we can ensure that our technique works well across different types of surfaces. One of the most important metrics of a thermometer is its temperature resolution. **Figure 4d** illustrates the plot of ratio temperature uncertainty versus integration time. As ratio vs temperature maintains a linear fit equation, the temperature uncertainty ($\sigma_T$) is calculated from $\sigma_R/m$ where $\sigma_R$ is the standard deviation of the ratio at a specific integration time and m is the slope of the calibration curve plotted in **Figure 2c**. Temperature resolution $\eta_T$ can be calculated from the equation $\eta_T = \sigma_T\sqrt{t_m}$, where $t_m$ is the integration time. It is clear from the equation that the temperature uncertainty $\sigma_T$ is related to integration time $t_m$. The temperature uncertainty $\sigma_T$ data is well fitted with the fit function $\frac{1}{\sqrt{t_m}}$ which is depicted by the purple line in **Figure 4d**. From the fit, we have calculated the temperature resolution to be $\eta_T = 0.72 \text{ KHz}^{-\frac{1}{2}}$. The temperature resolution with respect to the power density of our method is $22.91 \text{ KHz}^{-\frac{1}{2}}\text{Wcm}^{-2}$, where 10 mW is the excitation power, and the fiber core diameter is 200 μm. To the best of our knowledge, such temperature resolution relative to the power density is the best value achieved so far for fiber-based optical thermometers described in the literature. The relative sensitivity of our current method is 0.47% $K^{-1}$ at 338 K, which is among the highest, yet the main advantage of our method is that the sensor is reconfigurable and flexible, overcoming the shortcomings of non-fiber-based thermometry methods.

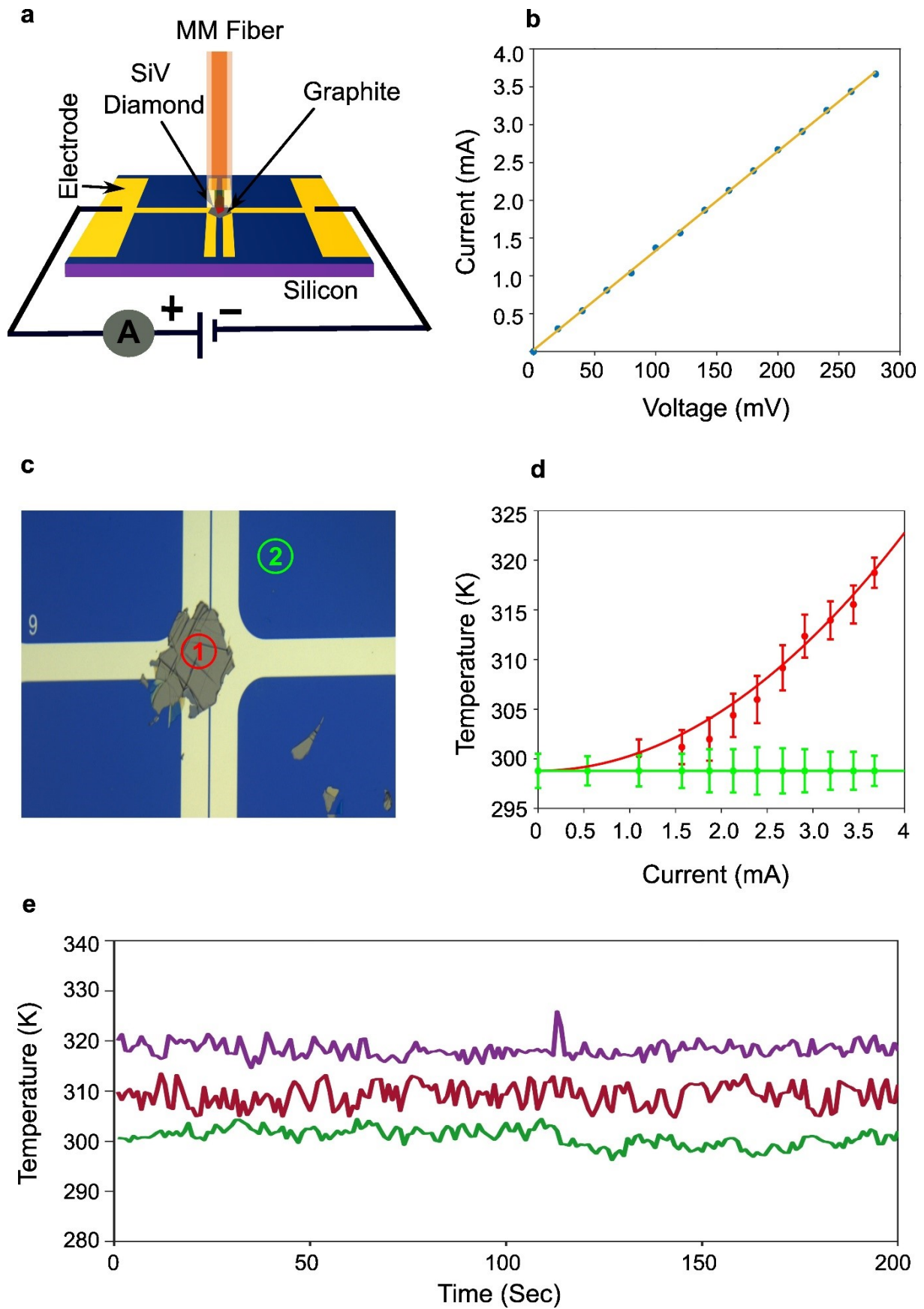

**Fig. 5** Real-time temperature measurement in a graphite heater device using our proposed fiber-based optical diamond thermometer. (a) Schematic of the fiber-based diamond

thermometer in contact with the graphite microheater, in which temperature values can be monitored. Local temperature is increased by applying an external voltage to the circuit. (b) I-V curve of the graphite heater device. The amber line is a linear fit to the experimental data which is marked by blue color. (c) Optical image of the graphite heater device. Two different locations are chosen in the device (red and green circles) for temperature measurement. (d) Real-time temperature readouts at two different locations of the device (red and green circles) are plotted with a function of current flowing through the graphite layer. Each temperature is calculated from the acquired SiV/laser count ratio data (**Supporting Information Figure S6**) with 1s integration time for 200s after the thermometer is calibrated against a known set of temperature and touching the known surface. The temperature data of the red circle actually fits with the simple quadratic equation $T(i) = i^2 a + T_0$, where a is the constant, $i$ is the current flowing through the graphite layer, and $T_0$ is the temperature offset. On the other hand, the temperature data of the green circle fits the straight-line equation which is plotted as the green line. The error bar of each data point is measured by taking the standard deviation of all the temperature data computed at a specific current. (e) Real-time temperature measurement on the graphite layer of the device as a function of time. Three distinct temperatures are measured for three different levels of current. Green, maroon, and purple lines represent temperature readings for 1.57 mA, 2.67 mA, and 3.67 mA current flowing through the graphite layer respectively. Each temperature readout is calculated from the SiV/laser ratio, with an integration time of 1 s. All the temperature data are taken with a 532 nm green laser with 10 mW excitation power.

Finally, our proposed fiber optic diamond thermometer is tested on a graphite-based heater device. The device is prepared by transferring graphite to a commercially available platinum-coated field-effect transistor test chip (**Cf. Methods**). **Figure 5a** illustrates the schematic of our sensor measuring the temperatures on the test device. **Supporting Information Figure S7** demonstrates the picture of a real-time temperature monitoring scenario on the test device. Graphite which is a highly conductive material is deposited on a 2 μm channel between the platinum electrode pair. We have used Raman spectroscopy to confirm that the transferred graphite is of high quality (**Supporting Information Figure S8**). External voltage is applied to the electrode pair to flow current through the graphite layer. Graphite is heated up quickly when current is flowing through it due to its power dissipation. **Figure 5b** shows the I-V characteristics curve of the microcircuit. Current increases from 0 mA to 3.67 mA when voltage is increased from 0 mV to 280 mV. The experimental data fits well with the linear equation, suggesting an ohmic behavior of the device. **Figure 5c** features the optical image of the device we used to demonstrate the sensing capabilities of the sensor. Location 1 (red circle) and Location 2 (green circle) are chosen to measure the temperature of the graphite and silicon dioxide part of the device while conducting the current. **Supporting Information Figure S6** shows the ratio versus the current plot for two different locations of the device. Every current value is maintained for ~2 minutes to ensure the thermal equilibrium is achieved, after which APD readings are taken. The ratio of Location 1 (on graphite) decreases with increasing current while the ratio of Location 2 (on silicon dioxide) remains constant. As the graphite surface and silicon dioxide surface give rise to different backscattered laser intensities to the fiber, different ratios are acquired at the same room temperature (298 K). To translate this ratio to the corresponding temperature we have done the offset of the ratio for both locations with respect to the calibration curve room temperature ratio. Then with the calibration curve linear fitting equation, the actual temperature reading is obtained as a function of the input current as shown in **Figure 5d.** As illustrated in the figure, the temperature exhibits minimal variation up to a current of 1.1 mA, but beyond that point, the device experiences a more rapid rise in temperature. Temperature increases from 298 K to 318 K by increasing current from 0 mA to 3.67 mA through the device. As the graphite layer is a resistive component, power dissipation through it can be calculated by the power equation, $P = i^2 R$. The temperature of the device also scales linearly with the power dissipation of the device.[36] The temperature versus current data fits well with the simple quadratic equation $T(i) = i^2 a + T_0$, which also agrees with our power dissipation equation of the device. On the other hand, the temperature of Location 2 remains constant while the current is increasing in the device. This is because Location 2 of the device, which is the silicon dioxide substrate of the device, is not conducting any current. As a result, there is no heat dissipation at that specific point of the device. Thus, local temperature monitoring is demonstrated in this microheater with our proposed thermosensor. Lastly, to confirm the real-time functionality of this thermometer, the recorded temperature as a function of time is illustrated in **Figure 5e**. With external voltage applied to the device, specific current is maintained for 200s, and temperature is recorded at every 1s utilizing the SiV/laser intensity ratio. As temperature is a function of current ($T(i) = i^2 a + T_0$), at specific current temperature remains almost constant over this 200s time period. The small deviation of temperature at specific current over the time is the result of the error bar of that respective temperature plotted in **Figure 5d.** We record temperature as a function of time for three different levels of current (1.57 mA, 2.67 mA, and 3.67 mA). As the current level is increased, the temperature also increases in accordance with the previously mentioned quadratic equation. Thus, real-time temperature monitoring is possible with our proposed thermometer.

## CONCLUSIONS

In conclusion, we have fabricated a fiber-based optical thermosensor based on SiV ensembles in microdiamond and proposed a new method to measure the temperature based on the photoluminescence intensity ratio of SiV and backscattered laser. A systematic and deterministic process of attaching the diamond to the fiber core is introduced in this work for the first time which will open new avenues in fiber-based thermometry research. Our thermosensor shows a good relative sensitivity of $0.47\%$ $K^{-1}$ at 338 K, spatial resolution of ~7.5 µm, and temperature resolution relative to power density of $22.91$ $\text{KHz}^{-\frac{1}{2}}\text{Wcm}^{-2}$, which is the best among all previously reported fiber-based all optical thermometers. The sensitivity can, of course, be improved by depositing more diamond emitters on the fiber core, but this comes at the expense of the spatial resolution of the sensor. The fiber-based optical thermometer is also characterized to have an optical collection efficiency of 5.5% compared to free-space collection without significant loss in its spectral properties. However, numerical analysis predicts that the system could still be optimized up to a maximum efficiency of 31.3% by improving the collection optics. We have also demonstrated the suitability of our thermosensor for real-time monitoring temperature on a microscale device such as the graphite-based microheater device in this work.

## METHODS

### Synthesis of SiV microdiamond

SiV microdiamond is synthesized with high pressure ~8 GPa and high temperature ~1800-2000 °C that is maintained for 1 minute in C-H-Si (0.19 atoms %) growth system. For the synthesis process, a powder mixture including 300 mg of Adamantane ($C_{10}H_{16}$) with a purity of over 99% from Sigma-Aldrich and 18 mg of Tetraphenylsilane ($C_{24}H_{20}Si$) with a purity of 96% from Aldrich is mixed for 5 minutes with a mortar and pestle that is made of jasper. Then, the mixture is pressed into a pellet weighing 65 mg and inserted into a titanium capsule. The diameter of the capsule is 6 mm with 0.2 mm wall thickness and has a height of 4 mm. In the reaction cell, to generate high pressure and temperature a toroid-type high-pressure chamber is used. Finally, the sample is rapidly cooled to room temperature under pressure. The microdiamond is then dispersed in IPA at a concentration of 0.1% (weight/weight) and stored in a small vial.

**Transferring diamond to fiber core**

The selected diamond dimension is measured using an atomic force microscope (AFM, Park XE7) before transferring it onto the fiber core (Figure 2c). A multimode fiber with a core diameter of 200 μm (Thorlabs M136L03) is used as the target fiber for transferring SiV diamond. The process of deterministic transfer of the diamond to the target fiber core is done by our lab-built transfer setup. This includes a microscope with a long working distance objective, a digital camera with LED, an X-Y micro positioner sample stage, and an X-Y-Z micro positioner stamping stage. Fiber is mounted on the sample stage by a standard fiber chuck (Thorlabs HFC005). The fiber chuck helps to hold the fiber securely. A Glass stamp (with PMMA/SiV diamond) is mounted on the stamping stage which is slowly lowered down with a Z micro-positioner. The optical microscope helps to align the SiV diamond to the fiber core precisely which is continuously monitored on the computer display by a digital camera.

**Fabrication of graphite-based heater device**

At first, a glass stamp containing a poly(dimethylsiloxane) (PDMS) dome is prepared. Two components of PDMS- the base and the curing agent are mixed in a weight ratio of 10:1 which is the recommended ratio and kept in a small vial. The volume of the base and the curing agent was 1000 μL and 100 μL, respectively. Then, the liquid mixing is de-aired in a vacuum box and kept in the box for 20 minutes to remove the air bubbles present in the mix. Then the mixture is cured at 60°C for 6 minutes. A small dimension of transparent glass (approx. 3 cm × 1 cm) is cut from a glass slide and placed on a hot plate which is set to 150°C. After that, using a needle a small droplet of mixing is dropped onto the small glass slide. That makes a small PDMS dome on the glass slide and the dome on the small glass is left on a hot plate for 5 minutes to finally cure it properly. After that, a PVA-based solution (Elmer's Liquid School Glue) with a concentration of 1:10 with respect to water is deposited on top of the PDMS dome. Next, we initiate the transfer of the graphite from commercially available tape (2D semiconductors) to the PDMS gel pak. Graphite flakes are mechanically exfoliated from the tape onto a PDMS substrate using the scotch tape method.[37] The PDMS/PVA is then used to pick up the graphite layers from the PDMS gel pak using our lab-built transfer stage. The transparency of the PDMS/PVA stamp helps us to deterministically pick up the graphite layers from the sample. After picking up it becomes ready to be deposited onto our target chip (Platinum OFET test chip, 2 μm channel width) which is treated beforehand by UV ozone cleaner. The target device is mounted on our sample stage and the stamp is mounted on the stamping stage of our transfer setup. The stamp with graphite layer is aligned to the target location (2 μm channel) of the device and slowly lowered down with Z micro-positioner until it makes contact with the target. After making contact, water is squirted into the contact point using a syringe. Then we wait for some time to dry the water and then slowly stamp is lifted, leaving the graphite layer onto the device. Then the device is heated at 60 °C with a hot plate for 2 minutes. Then PVA on top of the graphite layer is easily washed off with warm water and the device becomes ready to use. **Supporting Information Figure S8 illustrates** the Raman spectroscopy of the graphite after transferring it onto the chip. A peak at ~1582 cm$^{-1}$ is observed confirming the quality of graphite after transfer.

**Optical characterization**

Lab-built confocal microscopy setup is used to analyze the optical characteristics of the transferred SiV diamond attached to the multimode fiber core. A 532-nm continuous wave green laser (Cobolt Samba™ 532 nm) beam is focused on the SiV diamond on the fiber core through a long-working-distance objective with an NA of 0.7 (Thorlabs, MY100X-806, 100×) and the signal is collected by the spectrometer (ANDOR-SR-500i). The laser excitation signal and the emitted signal are split by a cube beam splitter 30R/70T (Thorlabs). For the confocal map and PL intensity (**Figure 2e** and **Figure 2f**) of the SiV diamond on the fiber 300 μW excitation power is used. To generate a confocal map, a scanning mirror (Newport, FSM-CD300B) is used to control the position of the laser spot position and scan across the fiber. The collection efficiency of the optical fiber with respect to the free-space collection via the objective is then investigated by determining the portion of the SiV diamond PL emission collected through the apertures of the fiber and the objective, respectively. The same optical setup for the confocal microscopy is used except that the opposite end of the fiber is directed to the spectrometer via a flip mirror connected to the optical path of the objective-based collection. Light exiting from the fiber is first collimated by an achromatic doublet lens (Thorlabs AC127-019-A) and guided by a pair of planar mirrors toward the flip mirror. With this, an alternating measurement between the objective-based and fiber-based collection is enabled in a single optical table. Once the collection efficiency of the fiber is investigated, we then move to set up as illustrated in Figure 1. Here

SiV diamond on the fiber bare end is excited by a 532 nm laser through the other end of the fiber and the signal is also collected from the end of the fiber. The flat cleaved fiber end coupled with SiV diamond is attached to a high-precision temperature controller (Microoptik-MDTC600) with a temperature resolution of 0.1° C. A plate beam splitter (30R/70T) (Thorlabs) is used to excite and collect signals from the same fiber. Reflected light excites the laser and transmitted light passes through the beam splitter, guided by a set of planar mirrors, and then again hits a plate beam splitter (50R:50T) (Thorlabs). The collection arm is then further divided into two paths, one for the SiV emission and the other for the laser emission. To maximize the SiV signal in one collection path a long pass filter (Semrock, FF01-715/LP-25) and a bandpass filter (Semrock, FF01-747/33-25) are placed. On the other hand, in the laser collection path, a short pass filter (Semrock, SP-01-532RU-25) and a neutral density filter (Semrock, ND2) are used in the laser collection arm. These filters limit the laser count for safe APD operation. The collection paths are fiber-coupled into two graded-index multimode fibers (Thorlabs, GIF625) which are connected to two separate single-photon avalanche photodiode (SPAPD) (Excelitas Technologies, SPCM-AQRH 14-FC) that counts all the incoming photons. LabVIEW software is developed to control all the hardware, analyze the photon rates, take the direct ratio of the SiV to laser emission intensity, and output the real-time fluorescence ratio based on the individual signals collected at each SPAPD. Alternatively, the photoluminescence from individual microdiamond is spectrally analyzed with a spectrometer (ANDOR-SR-500i) equipped with a charge-coupled device (CCD) camera.

**Numerical Simulation**

To provide theoretical insight into the collection efficiency of SiV emission to the optical fiber and objective, a 3D semi-analytic ray-optical analysis is performed which takes into account the far-field radiation pattern of the dipole emitters and the acceptance angles of the optical fiber and objective. Due to the dipole nature of the emitters, the far-field radiation pattern $R(r, \theta, \varphi)$ is modeled as equation N in spherical coordinates over $\theta \in [0, \pi]$ and $\varphi \in [0, 2\pi]$ with respect to the dipole axis ($\theta = 0$).

$$R(r, \theta, \varphi) \propto sin^2(\theta)$$

Further calculation is then carried out in the Cartesian coordinate system. The crystallographic facet <111> is set to correspond to the XY plane, so that the Case I dipole becomes parallel to the z-axis and its positive end is set to the origin, for simplicity. The radiation patterns for Cases II, III, and IV are then obtained by performing appropriate rotation operations to the Case I dipole. **Supporting Information Equation S1** shows the rotation matrix operators used while **Supporting Information Table S2** shows the mathematical expressions to calculate the coordinates of each dipole orientation. For each dipole case, the intersection of the radiation pattern to the acceptance cone defined by the acceptance angle ($\theta_{ACC}$) of the fiber and the objective centered along the z-axis is calculated. The sum of all radial distances within the intersection is then compared to the sum of all radial distances of the radiation pattern to give the collection efficiency ($\eta_i, i = I, II, III, IV$) with respect to the total emission.

$$\eta_i = \frac{P_{Acc,i}}{P_{Total,i}} = \frac{\sum_{\phi}^{2\pi}\sum_{\theta}^{\theta_{Acc}} R_i(r,\theta,\varphi)}{\sum_{\phi}^{2\pi}\sum_{\theta}^{\pi} R_i(r,\theta,\varphi)}$$

This is performed on all dipole orientations and the sum is calculated to provide the overall collection efficiency (η) for the SiV diamond emitter system.

$$\eta = \frac{\sum_{i=I}^{i=IV} P_{Acc,i}}{\sum_{i=I}^{i=IV} P_{Total,i}}$$

The relative collection efficiency of the fiber to the objective ($\frac{\eta_{Fiber}}{\eta_{Objective}}$) is then evaluated to set the theoretical upper limit for this collection system and is compared to the intensities of the experimental spectral measurements.

## ASSOCIATED CONTENT
### Data Availability Statement

The datasets generated during and/or analyzed during the current study are available from the corresponding author upon reasonable request.

## AUTHOR INFORMATION
### Author contribution

S. H. and T. T. T. conceived the idea of the project and designed the experiments. S. H., Y. C. and T. T. T built the optical system and its software. E. E. fabricated SiV microdiamonds. S. H. fabricated the fiber-based SiV thermal sensor. G. P. and S. H. built the graphite microcircuit. S. H. performed all the optical characterization and the thermometric measurements. M. B. and A. S. conducted the semi-analytical modeling of the SiV emission.

T. N. A. M. conducted the AFM measurement. S. H. and T. T. T. analyzed the data. T. T. T. supervised the project. All authors discussed the results and co-wrote the manuscript.

## Acknowledgment

S.H. is grateful to UTS for the support to this research through the International Research Scholarship scheme. The authors thank the UTS node of Optofab ANFF for the assistance with nanofabrication.

## Funding Sources

T. T. T. and S.H. acknowledge the Australian Research Council (DE220100487) for financial support. E. E. is grateful to RFBR and GACR for the support, project number 20-52-26017.

## Notes

The authors declare no competing financial interest.

# Supporting information

# Fiber-based Ratiometric Optical Thermometry with Silicon-Vacancy in Microdiamonds

Md Shakhawath Hossain,[†] Miguel Bacaoco,[‡] Thi Ngoc Anh Mai,[†] Guillaume Ponchon,[§] Chaohao Chen,[∥,#] Lei Ding,[††] Yongliang Chen,[‡‡] Evgeny Ekimov,[§§,∥∥] Helen Xu,[††] Alexander S. Solntsev,[‡] and Toan Trong Tran [†,*]

[†]School of Electrical and Data Engineering, University of Technology Sydney, Ultimo, NSW, 2007, Australia

[‡]School of Mathematical and Physical Sciences, University of Technology Sydney, Ultimo, NSW, 2007, Australia

§École polytechnique universitaire de l'université Paris-Saclay (Polytech Paris-Saclay), Université Paris-Saclay, Bâtiment 620, Maison de l'Ingénieur, Rue Louis de Broglie, 91190 Orsay.

‖Department of Electronic Materials Engineering, Research School of Physics, The Australian National University, Canberra, Australian Capital Territory 2601, Australia

#ARC Centre of Excellence for Transformative Meta-Optical Systems (TMOS), Research School of Physics, The Australian National University, Canberra, Australian Capital Territory 2601, Australia

††School of Biomedical Engineering, University of Technology Sydney, Ultimo, NSW, 2007, Australia

‡‡ Department of Physics, The University of Hong Kong, Pokfulam, Hong Kong, China.

§§Institute for High Pressure Physics, Russian Academy of Sciences, Troitsk 142190, Russia

‖‖Lebedev Physics Institute, Russian Academy of Sciences, Moscow 117924, Russia

*Corresponding author: trongtoan.tran@uts.edu.au

The Supporting Information includes:

**Supporting Information Figure S1:** Raman spectra of SiV microdiamond deposited on fiber core.

**Supporting Information Figure S2:** Avalanche photodiode (APD) counts per second obtained from the SiV collection path and laser collection path.

**Supporting Information Figure S3:** Repeatability test of SiV/laser ratio vs temperature plot acquired from four back-to-back heat cycles.

**Supporting Information Figure S4:** SiV/laser ratio as a function of excitation power.

**Supporting Information Figure S5:** SiV/laser ratio versus temperature plot when the fiber sensor is 1cm away from the heating stage of the temperature controller.

**Supporting Information Figure S6:** SiV/laser ratio versus current plot of the graphite microheater used in this work.

**Supporting Information Figure S7** Photograph of the real-time temperature monitoring of the graphite microheater.

**Supporting Information Figure S8:** Raman spectra of the graphite layer on the OFET test chip.

**Supporting Information Table S1.** The calculated collection efficiency of the optical fiber and objective per dipole case.

**Supporting Information Table S2.** Expressions for the coordinates of each dipole radiation from spherical to Cartesian and its subsequent rotations.

**Supporting Information Equation S1.** Rotation matrix operators along the x-axis ($r_x$) and z-axis ($r_z$) axis used to calculate the coordinates of the dipole radiation of oblique cases (Cases II, III, IV) from Case I.

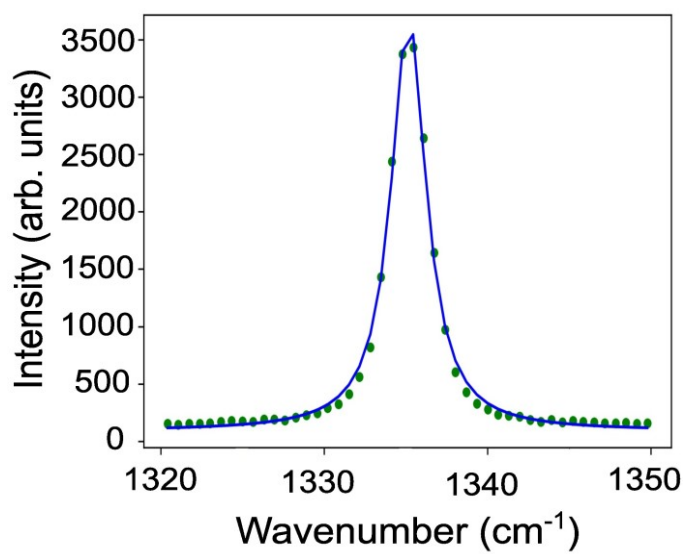

**Figure S1** Raman spectra of the corresponding SiV microdiamond on the fiber core, fitted with a Lorentzian function. Raman peak is at ~1335 cm$^{-1}$. The spectrum is taken with 30s acquisition time and 1.7 mW excitation power through the objective.

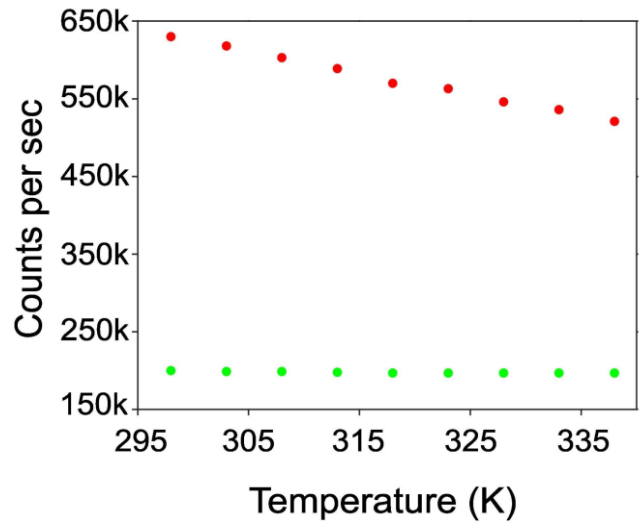

**Figure S2** Avalanche photodiode (APD) counts per second obtained from the SiV (red circles) collection path and laser (green circles) collection path (cf. main text, Figure 4).

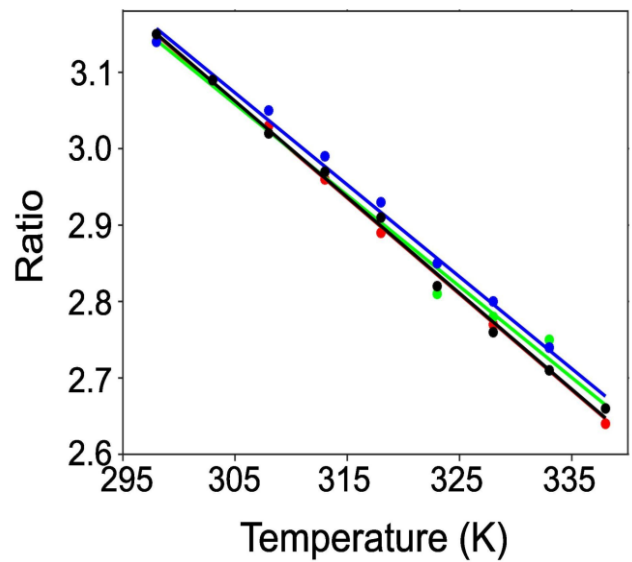

**Figure S3** Repeatability test (cf. main text, Figure 4). The SiV/laser ratio vs. temperature plot is acquired from four back-to-back heat cycles. Among four cycles, two are heating (red, blue circles) and two are cooling (green, black circles) sweeps. The four linear fitting equations ($f(T) = -mT + C$) are applied to the raw data and color-coded accordingly.

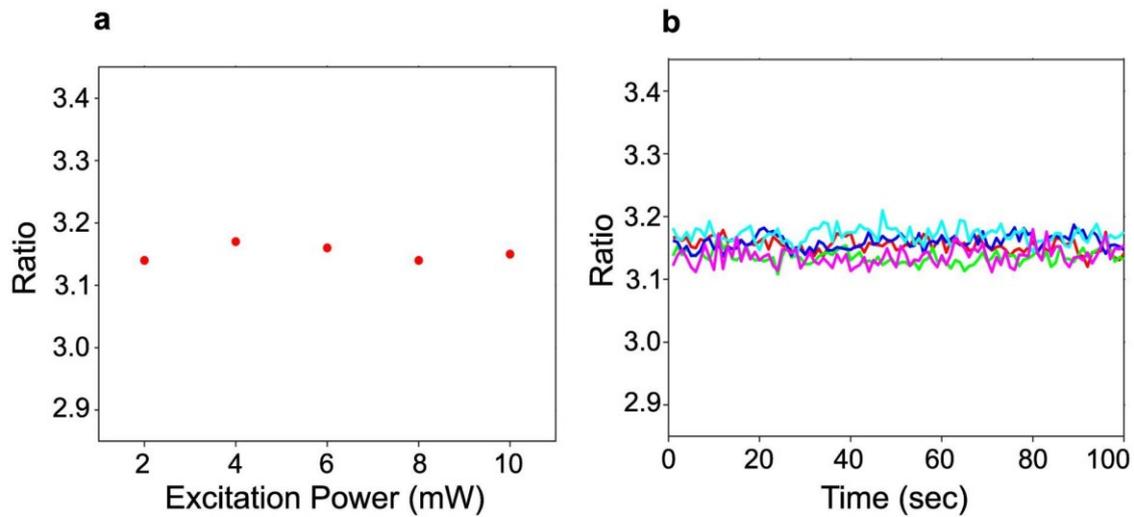

**Figure S4** (a) SiV/laser ratio as a function of excitation power. The ratio remains almost constant at different excitation power. (b) The SiV/laser emission ratio is plotted as a function of time for various excitation powers as shown in (a).

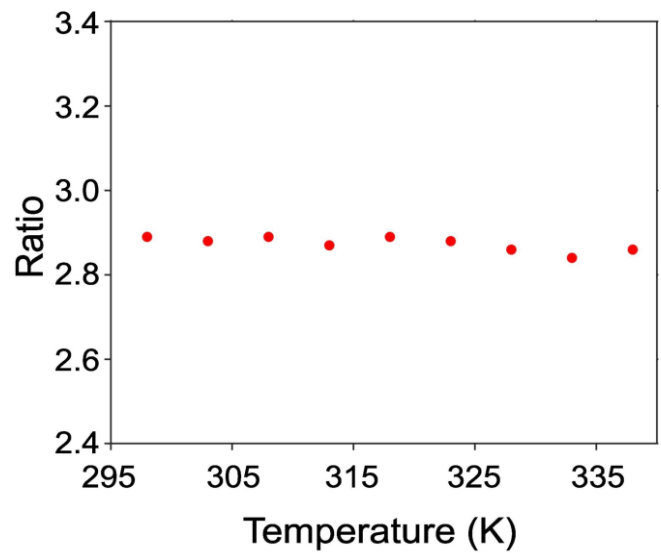

**Figure S5** SiV/laser ratio versus temperature plot when the fiber sensor is 1cm away from the heating stage of the temperature controller. The ratio remains invariant to the increase in temperature, suggesting that surface reflectivity does not change with temperature in our experiments.

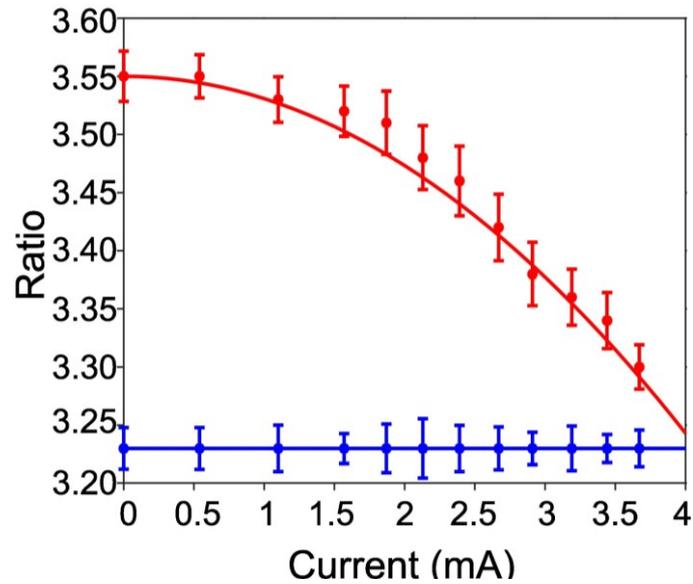

**Figure S6** SiV/laser ratio versus current plot of the graphite microheater used in this work (cf. main text, Figure 5). The red circle represents the ratio when the fiber touches the graphite layer and the blue circle represents the data when the fiber touches the silicon layer on the device. The red line is a quadratic equation $R(i) = -i^2 a + b$ fit and the blue line is a linear fit to the experimental data.

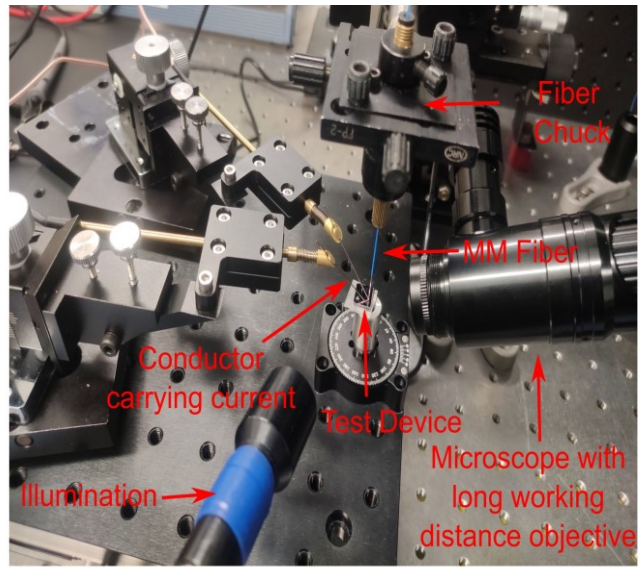

**Figure S7** Photograph of the real-time temperature monitoring of the graphite-based microheater.

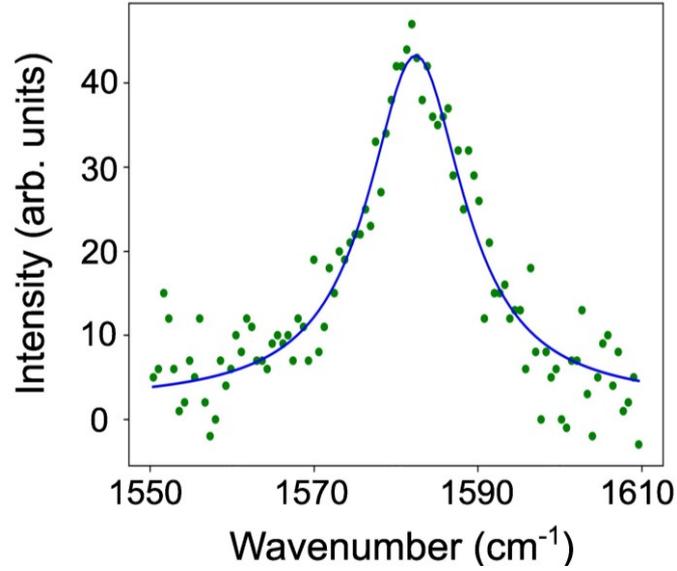

**Figure S8** Raman spectra of the graphite layer on the OFET test chip, fitted with a Lorentzian function. Raman peak is at ~1582 cm$^{-1}$. The spectrum is taken with 30s acquisition time and 1.7 mW excitation power through the objective.

**Table S1.** The calculated collection efficiency of the optical fiber and objective per dipole case.

|  | Case I | Case II | Case III | Case IV | Sum |
|---|---|---|---|---|---|
| Optical Fiber collection efficiency (%) | 0.095 | 0.824 | 0.824 | 0.824 | 0.642 |
| Objective collection efficiency (%) | 0.530 | 2.558 | 2.558 | 2.558 | 2.051 |
| Fiber to objective efficiency (%) | 17.913 | 32.218 | 32.218 | 32.218 | 31.294 |

**Equation S1.** Rotation matrix operators along the x-axis ($r_x$) and z-axis ($r_z$) axis used to calculate the coordinates of the dipole radiation of oblique cases (Cases II, III, IV) from Case I.

$$r_x(\alpha) = \begin{vmatrix} 1 & 0 & 0 \\ 0 & \cos(\alpha) & -\sin(\alpha) \\ 0 & \sin(\alpha) & \cos(\alpha) \end{vmatrix}$$

$$r_z(\theta) = \begin{vmatrix} \cos(\theta) & -\sin(\theta) & 0 \\ \sin(\theta) & \cos(\theta) & 0 \\ 0 & 0 & 1 \end{vmatrix}$$

**Table S2.** Expressions for the coordinates of each dipole radiation from spherical to Cartesian and its subsequent rotations.

| Dipole Case | Cartesian Coordinates |
| --- | --- |
| I | $[x1, y1, z1] = [R\sin(\theta)\cos(\varphi), R\sin(\theta)\cos(\varphi)\, R\cos(\theta)]$ |
| II | $[x1, y1, z1] * r_x(\alpha = 70.5)$ |
| III | $[x1, y1, z1] * r_x(\alpha = 70.5) * r_z(\theta = 120)$ |
| IV | $[x1, y1, z1] * r_x(\alpha = 70.5) * r_z(\theta = -120)$ |